\documentclass[12pt,english]{paper}
\usepackage{amssymb}
\usepackage[T1]{fontenc}
\usepackage[latin1]{inputenc}
\usepackage{graphicx}
\usepackage{cite}
\makeatletter

\newcommand{\bibstyle@aas}{\bibpunct{(}{)}{;}{a}{}{,}}

\makeatletter

\makeatletter

\usepackage{geometry}

\makeatletter

\usepackage{epsfig}

\makeatother

\makeatother

\makeatother

\usepackage{babel}
\makeatother
\begin{document}

\title{Hints on the Existence of Superheavy Dark matter and Ultrahigh Energy Dark Matter in Extreme Energy Cosmic Rays Observations}

\author{Ye Xu$^{1,2}$}

\maketitle

\begin{flushleft}
$^1$School of Information Science and Engineering,  Fujian University of Technology, Fuzhou 350118, China
\par
$^2$Research center for Microelectronics Technology, Fujian University of Technology, Fuzhou 350118, China
\par
e-mail address: xuy@fjut.edu.cn
\end{flushleft}

\begin{abstract}
In the present paper, it is assumed that there exist two dark matter particles: superheavy dark matter particles (SHDM), whose mass $\sim$ inflaton mass, and lighter fermion dark matter particles (LFDM) which are the ultra-relativistic products of its decay. The event rates of LFDM measured by the Pierre Auger observatory (Auger) are evaluated in the energy range between 1 EeV and 1 ZeV at the zenith angles between 0$^{\circ}$ and 60$^{\circ}$ when the different lifetimes of decay of SHDM. Thus it is proved that the possibility of measurement of these LFDM at Auger. Based on the assumption that the detection efficiency for a shower caused by LFDM is set to be 100\%, the LFDM contribution to the energy spectrum of extreme energy cosmic rays (EECRs) is discussed. And it is found that there are some hints on the existence of SHDM and LFDM in the past EECRs observations.
\end{abstract}


\section{Introduction}
The nature and origin of dark matter (DM) remains one of the unanswered puzzles in particle physics, cosmology and astrophysics. But sufficient evidences for the existence of DM and its dominance in matter in our universe are provided by cosmological and astrophysical observations\cite{bergstrom,BHS}. It is indicated by the Planck observations with measurements of the cosmic microwave background that $26.6\%$ of the overall energy density of the universe is nonbaryonic DM\cite{Planck2015}. A restriction on dark matter, which indicate that the DM particle is nonrelativistic or "cold", arises from large-scale N-body simulations\cite{NWF,springel}. The DM mass is constrained by Local Thermal Equilibrium (LTE) in the early Universe and the partial wave unitarity of the S matrix (1 keV $\lesssim m_{DM}\lesssim$ 100 TeV\cite{VLHMR,GK}). Weakly Interacting Massive Particles (WIMPs), predicted by extensions of the Standard Model (SM) of particle physics, are a class of candidates for dark matter\cite{BHS}. They are distributed in a halo surrounding a galaxy. This WIMP halo with a local density of 0.3 GeV/cm$^3$ is assumed and its relative speed to the Sun is 230 km/s\cite{JP}. At present, one mainly searches for WIMPs via direct and indirect detections\cite{CDMSII,CDEX,XENON1T,LUX,PANDAX,AMS-02,DAMPE,fermi}. Because of the very small cross sections of the interactions between these WIMPs and nuclei (maybe O(10$^{-47}$ cm$^2$))\cite{XENON1T,PANDAX}, so far no one has found this thermal DM yet.
\par
The Superheavy Dark Matter (SHDM) $\phi$, based on the possibility of particle production due to time varying gravitational fields, is an alternative DM scenario\cite{KC87,CKR98,CKR99,KT,KCR,CKRT,CCKR,KST,CGIT,FKMY,FKM}. This scenario is based on three hypotheses: (a) SHDM in the early Universe never reaches LTE; (b) SHDM has a mass with the same order of magnitude as the inflaton mass; (c) the lifetime of SHDM exceeds the age of the Universe, that is $\tau_{\phi} \gg t_0$\cite{AMO,EIP}. Since direct detection of SHDM is unattainable, the best way to measure SHDM is through the indirect detection of its decay or annihilation products. The annihilation of SHDM particles is unobservable since their cross section is bounded by unitarity ($\sigma_{ann} \propto \displaystyle\frac{1}{m_{\phi}^2}$). So searching for the products of decay of SHDM is an unique way to test the existence of SHDM. One has considered the decay of SHDM as a source of very high energy or ultrahigh energy (UHE) cosmic rays in the past\cite{FG,EGS,hill,EGL,BKV,ST,BD,ABK,EIP,BLS,AMO}. This is the so-callded "top-down" models, where a flux of protons, gamma rays and neutrinos is produced by the decay of SHDM into the partons. Conversely, it is different from this decay mode referred to above that the products of the decay of SHDM are a class of lighter fermion dark matter (LFDM)\cite{BGG,BGGM,xu1,xu2,xu3}, not SM particles, in the present work.
\par
Then it is a reasonable assumption that there exist two DM species in the Universe. One is a non-thermal and non-relativistic dark sector generated by the early universe with its bulk comprised of a super massive relic, SHDM $\phi$ ($m_{\phi} \sim m_{inflaton}$), in the Universe. The other is the stable LFDM ($\chi$) which is UHE products of the decay of SHDM ($\phi\to\chi\bar{\chi}$). It is assumed that SHDM comprises of the bulk of present-day DM. Since the decay of long-living SHDM, meanwhile, the present-day DM may also contain a small component which is UHE LFDM. Although the fraction of these relativistic LFDM particles is small in the Universe, their large interaction cross sections (including between themselves and between them and SM particles) make it possible to measure them. In the present paper, a Z$^{\prime}$ portal dark matter model\cite{APQ,Hooper} is taken for LFDM $\chi$ to interact with nuclei. And, for the $\chi\chi$Z$^{\prime}$ and qqZ$^{\prime}$ interactions, their vertexes are both assumed to be vector-like. These UHE LFDM particles may be directly measured by their interaction with nuclei. Thus it is indicated that there exist SHDM particles in the Universe.
\par
Extreme energy cosmic rays (EECRs) is a puzzling and important topic in astropaticle physics, especially, their energy spectrum and components. There is the GZK-cutoff in the energy spectrum of EECRs at about $5\times10^{19}$ eV since extreme energy cosmic protons interact with photons for the Cosmic microwave background (CMB). Although they have been measured by some EECRs observatories in the past\cite{AGASA,HiRes,Auger2015,TA}, it is difficult to determine their spectrum and components because of the lack of sufficient exposure at energies above $10^{19}$ eV. But these observations of EECRs in the past may be explained by the SHDM scenario mentioned above. In what follows , it will be discussed there are some hints on the existence of UHE LFDM due to the decay of SHDM in the observations from Pierre Auger observatory (Auger) and Telescope Array (TA) near the GZK-off scale. Besides, since the diffuse astrophysical neutrinos may contribute to the EECRs spectrum, they will be discussed in what follows, too.
\par
\section{Estimation of UHE LFDM flux}
It is considered a scenario where the dark matter sector is composed of two particle species in the Universe. One is a non-thermal and non-relativistic particle species $\phi$, with mass $m_{\phi} \geq$ 1 EeV, the other is much lighter particle species $\chi$ ($m_{\chi} \ll m_{\phi}$), due to the decay of $\phi$, with a very large lifetime. And $\phi$ comprises the bulk of present-day dark matter. The lifetime for the decay of SHDM to SM particles is strongly constrained ($\tau \geq$ O($10^{26}-10^{29}$)s) by diffuse gamma and neutrino observations\cite{EIP,MB,RKP,KKK}. In the present work, it is considered an assumption that SHDM could decay to LFDM, not SM particles. So $\tau_{\phi}$ is taken to be between $10^{17}$ s (the age of the Universe) and $10^{26}$ s.
\par
The LFDM flux is composed of galactic and extragalactic components. So the total flux $\psi_{\chi}=\psi_{\chi}^G+\psi_{\chi}^{EG}$, where $\psi_{\chi}^G$ and $\psi_{\chi}^{EG}$ are the LFDM galactic and extragalactic fluxes, respectively.  The LFDM flux from the Galaxy only depends on the two-body decay of SHDM particles and their distribution in the galactic halo, it is the same as the neutrino flux due to the decay of SHDM in Ref.\cite{EIP,BLS}:
\begin{center}
\begin{equation}
\psi_{\chi}=\displaystyle\frac{1}{4\pi m_{\phi}\tau_{\phi}}\int \displaystyle\frac{dN_{\chi}}{dE_{\chi}}\rho_{halo}dsdE
\end{equation}
\end{center}
where $\rho_{halo}$ is the density profile of dark matter particles in the Galaxy and s is a line-of-sight. $\displaystyle\frac{dN_{\chi}}{dE_{\chi}}=2\delta(E_{\chi}-\displaystyle\frac{m_{\phi}}{2})$, and E$_{\chi}$ and N$_{\chi}$ are the energy and number of UR LFDM, respectively.
Then the UHE LFDM flux from the Galaxy is obtained via the following equation\cite{BLS}:
\begin{center}
\begin{equation}
\psi_{\chi}=\int_{E_{min}}^{E_{max}}F^G\frac{dN_\chi}{dE_\chi}dE
\end{equation}
\end{center}
with
\par
\begin{center}
\begin{equation}
F^G=1.7\times10^{-8}\times\frac{10^{26}s}{\tau_{\phi}}\times\frac{1TeV}{m_{\phi}} cm^{-2}s^{-1}sr^{-1}.
\end{equation}
\end{center}
\par
The UHE LFDM flux from the extra galaxy is obtained via the following equation\cite{BGG,EIP}:
\begin{center}
\begin{equation}
\psi_{\chi}^{EG}=F^{EG}\int_{E_{min}}^{E_{max}}dE \int_0^{\infty}dz\frac{1}{\sqrt{\Omega_{\Lambda}+\Omega_m(1+z)^3}}\frac{dN_\chi}{dE_\chi}[(1+z)E_\chi]
\end{equation}
\end{center}
with
\par
\begin{center}
\begin{equation}
F^{EG}=1.4\times10^{-8}\times\frac{10^{26}s}{\tau_{\phi}}\times\frac{1TeV}{m_{\phi}} cm^{-2}s^{-1}sr^{-1}.
\end{equation}
\end{center}
where z represents the red-shift of the source, $\Omega_{\Lambda}=0.685$ and $\Omega_m=0.315$ from the PLANCK experiment\cite{Planck2015}. $\displaystyle\frac{dN_{\chi}}{dE_{\chi}}=2\delta(E_{\chi}-\displaystyle\frac{m_{\phi}}{2})$, where E$_{\chi}$ and N$_{\chi}$ are the energy and number of UHE LFDM, respectively.
\section{UHE LFDM and neutrino interactions with nuclei}
In the present paper, a $Z^{\prime}$ portal dark matter model\cite{APQ,Hooper} is taken for LFDM to interact with nuclei via a neutral current interaction mediated by a heavy gauge boson $Z^{\prime}$. This new gauge boson is considered as a simple and well-motivated extension of SM (see Fig.1(a) in Ref.\cite{BGG}). Since the interaction vertexes ($\chi\chi Z^{\prime}$ and $qqZ^{\prime}$) are assumed to be vector-like in the present work, the effective interaction Lagrangian can be written as follows:
\begin{center}
\begin{equation}
\mathcal{L} = \bar{\chi}g_{\chi\chi Z^{\prime}}\gamma^{\mu}\chi Z^{\prime}_{\mu} + \sum_{q_i} \bar{q_i}g_{qqZ^{\prime}}\gamma^{\mu}q_iZ^{\prime}_{\mu}
\end{equation}
\end{center}
where $q_i$'s are denoting the SM quarks, and $g_{\chi\chi Z^{\prime}}$ and $g_{qqZ^{\prime}}$ are denoting the $Z^{\prime}$-$\chi$ and $Z^{\prime}$-$q_i$ couplings, respectively.
This Deep inelastic scattering (DIS) cross-section is computed in the lab-frame and its parameters are taken to be the same as the ones in Ref.\cite{BGG}, that is, the coupling constant G ($G=g_{\chi\chi Z^{\prime}}g_{qqZ^{\prime}}$) is chosen to be 0.05 and the $Z^{\prime}$ and $\chi$ masses are taken to be 5 TeV, 10 GeV, respectively. Theoretical models that encompass the LFDM spectrum have been discussed in the literature in terms of Z or $Z^{\prime}$ portal sectors with $Z^{\prime}$ vector boson typically acquiring mass through the breaking of an additional U(1) gauge group at the high energies (see Ref.\cite{APQ,Hooper}). The DIS cross section for LFDM interaction with nuclei is obtained by the following function(see Fig.1(b) in Ref.\cite{BGG}):
\begin{center}
\begin{equation}
\sigma_{\chi N}=6.13\times10^{-43} cm^2 \left(\frac{E_{\chi}}{1GeV}\right)^{0.518}
\end{equation}
\end{center}
where E$_{\chi}$ is the UHE LFDM energy.
\par
The DIS cross-sction for UHE neutrino interaction with nuclei is computed in the lab-frame and given by simple power-law forms\cite{BHM} for neutrino energies above 1 EeV:
\begin{center}
\begin{equation}
\sigma_{\nu N}(CC)=4.74\times10^{-35} cm^2 \left(\frac{E_{\nu}}{1 GeV}\right)^{0.251}
\end{equation}
\end{center}
\begin{center}
\begin{equation}
\sigma_{\nu N}(NC)=1.80\times10^{-35} cm^2 \left(\frac{E_{\nu}}{1 GeV}\right)^{0.256}
\end{equation}
\end{center}
where $E_{\nu}$ is the neutrino energy. Then the above equations show that $\sigma_{\chi N}$ is smaller by 10-11 orders of magnitude, compared to $\sigma_{\nu,\bar{\nu} N}$, near the GZK-cutoff scale.
\par
The LFDM and neutrino interaction lengths can be obtained by
\par
\begin{center}
\begin{equation}
L_{\nu,\chi}=\frac{1}{N_A\rho\sigma_{\nu,\chi N}}
\end{equation}
\end{center}
\par
where $N_A$ is the Avogadro constant, and $\rho$ is the density of matter, which LFDM and neutrinos interact with.
\section{Evaluation of the numbers of UHE LFDM and neutrinos measured by Auger}
Auger is a detector array, covering an area of 3000 km$^2$, for the measurement of EECRs and located outside the town of Malargue, in the Province of Mendoza, Argentina. EECRs are detected in a hybrid mode at Auger, that is it consists of about 1600 surface detectors (SD) to measure secondary particles at ground level and four fluorescence detectors (FD), each consisting of 6 optical telescopes, to measure the development of extensive air showers (EAS) in the atmosphere\cite{Auger2010}. The secondary hadrons are produced by UHE LFDM (or neutrinos) reaching the Earth via the LFDM-nucleus (or neutrino-nucleus) DIS. These secondary particles will develop into an EAS measured by SD and FD at Auger. In the present paper, it is made an assumption that there exists air under an altitude of H = 100 km.
\par
The number of UHE LFDM, N$_{det}$, detected by Auger can be obtained by the following function:
\begin{center}
\begin{equation}
N_{det} = R\times T\times \int^{E_{max}}_{E_{min}} \int_{\Omega} \int_{S_{eff}} \eta \Phi_{\chi} P(E,D(\theta))cos(\theta) dS d\Omega dE
\end{equation}
\end{center}
where $\Omega$ is the solid angle and $\theta$ is the zenith angles between 0$^{\circ}$ and 60$^{\circ}$. R is the duty cycle for Auger and taken to be 100\%. T is the lifetime of taking data for Auger and taken to be 10 years in the present work. dS=dx$\times$dy is the horizontal surface element. $S_{eff}$ is the effective observational area for Auger and about 3000 km$^2$. E is the energy of an incoming particle and varies from $E_{min}$ to $E_{max}$. $\Phi_\chi=\displaystyle\frac{d\psi_\chi}{dE_{\chi}}$ and $P(E,D(\theta))=1-exp(-\displaystyle\frac{D}{L_{air}})$. $L_{air}$ is the UHE LFDM interaction lengths with the air and D($\theta$) is the effective length in the atmosphere above the Auger array and $D(\theta) = \displaystyle\frac{H}{cos(\theta)}$. Since $N_{det}$ is just Roughly estimated in the present paper, the efficiency for measuring a shower produced by UHE LFDM or neutrino $\eta$ is assumed to be 100\%.
\par
The diffuse astrophysical neutrinos is roughly estimated with a diffuse neutrino flux of $\Phi_{\nu}=0.9^{+0.30}_{-0.27}\times(E_{\nu}/100 TeV)\times10^{-18} GeV^{-1} cm^{-2}s^{-1}sr^{-1}$\cite{icecube}, where $\Phi_{\nu}$ represents the per-flavor flux, by the above method.

\section{Results}
The numbers of UHE LFDM and neutrino detected by Auger are evaluated at different energies (1 EeV < E <1 ZeV), respectively. Since Auger can only measure the deposited energy for an EAS, $E_{sh}$, in the atmosphere, it is important to determine the inelasticity parameter $y$. $y=1 - \displaystyle\frac{E_{\chi^{\prime},lepton}}{E_{in}}$ (where $E_{in}$ is the incoming LFDM or neutrino energy and $E_{\chi^{\prime},lepton}$ is the outgoing LFDM or lepton energy). For LFDM, $E_{sh}=yE_{in}$. Since an EAS due to the neutrino interaction with nuclei via a neutral current is much smaller than that via a charged current, the charged current be only considered in the neutrino interaction with nuclei. Then $E_{sh}=(1-y)E_{in}$ for neutrinos. The mean values of $y$ for LFDM and neutrinos are computed by Ref.\cite{BGG} and \cite{GQRS}, respectively, and their results are used to the calculation of $E_{sh}$ in the present paper. Fig. 1 shows the event rates for LFDM and astrophysical neutrinos measured by Auger at different $E_{sh}$'s. The numbers of the detected UHE LFDM can reach about 400 and 2 at the energies with 1 EeV and 1 ZeV in ten years when $\tau_{\phi} = 2\times10^{23}$ s, respectively, as shown in Fig. 1 (see the red black solid line). Since $\Phi_{\chi}$ is proportional to $\displaystyle\frac{1}{\tau_{\phi}}$, the above results are actually depended on the lifetime of SHDM. For example, the numbers of the detected UHE LFDM can reach about 50 and 8 at 1 ZeV in ten years when $\tau_{\phi} = 7\times10^{21}$ s (see the blue dot line), $4\times10^{22}$ s (see the red short dot line), respectively. The numbers of the detected UHE neutrinos can reach about 6 and $3.5\times10^{-5}$ at the energies with 1 EeV and 1 ZeV in ten years, respectively (see the purple dash line). The event rates for astrophysical neutrinos is smaller by 3-4 orders of magnitude at the GZK-cutoff scale, compared to LFDM when $\tau_{\phi} = 2\times10^{23}$ s, as shown in Fig. 1. So the astrophysical neutrino contribution to the energy spectrum of EECRs at the GZK-cutoff scale is neglected at all.
\section{Discussion and conclusion}
According to the results described above, it is possible that LFDM at the GZK-cutoff scale are directly detected at Auger when $\tau_{\phi} \lesssim 10^{23}$ s. Then these LFDM particles may have been measured by EECRs observatories and contribute the energy spectrum of EECRs, especially at the GZK-cutoff scale. The LFDM spectrum will be compared to the results of the EECRs observatories in the present work.
\par
Fig. 2 and 3 show that the LFDM spectra when $\tau_{\phi}$ = $7\times10^{21}$ and $4\times10^{22}$ s and the fluxes of EECRs measured by the Auger (including the ICRC 2015 Auger data\cite{Auger2015} and ICRC 2019 Auger data at ultrahigh energies\cite{Auger2019}) and TA\cite{TA}, respectively. In these figures, the ICRC 2015 Auger energy spectrum (shown with 'Best fit (Auger2015)') was obtained with the best-fit parameters for the SPG model (SimProp code, PBS photo-production cross-sections)\cite{ABGPS} in Ref.\cite{Auger}. The LFDM spectrum when $\tau_{\phi} = 4\times 10^{22}$ s (shown with 'LFDM') and total spectrum of it and the ICRC 2015 Auger energy spectrum ('LFDM+Best fit (Auger2015)') are shown in Fig. 2. This total spectrum is well consistent with the ICRC 2019 Auger data near $10^{20}$ eV. The LFDM spectrum when $\tau_{\phi} = 7\times 10^{21}$ s (shown with 'LFDM') and total spectrum of it and the ICRC 2015 Auger energy spectrum ('LFDM+Best fit (Auger2015)') are shown in Fig. 3. This total spectrum is consistent with the TA data near $10^{20}$ eV. These hints might reveal UHE LFDMs at near $10^{20}$ eV have been measured by the Auger or TA experiment in the past observations.
\par
Based on Fig. 2 and 3, it may be assumed that the ICRC 2015 Auger data at ultrahigh energies is well consistent with the GZK-cutoff and UHE LFDMs have been measured by the EECRs observatories. That is the energy spectrum of EECRs consists of two components: the proton GZK-cutoff and UHE LFDM spectrum and UHE LFDMs dominate the EECRs spectrum at energies above $10^{20}$ eV. The ICRC 2019 Auger data is more consistent with the LFDM spectrum when $\tau_{\phi} = 4\times 10^{22}$ s. The TA data is more consistent with the LFDM spectrum when $\tau_{\phi} = 7\times 10^{21}$ s.
\par
Certainly, the above results are based on the assumption that the efficiency for measuring a shower caused by UHE LFDM or neutrino is set to be 100\%. Besides, sufficient exposure will be used to determine the EECRs spectrum and its component in EECRs observatories. The JEM-EUSO Telescope is of the larger observational area ($2\times10^5$ km$^2$ in the Nadir mode and $7\times10^5$ km$^2$ in the tilted mode, its duty cycle $\sim$ 10-20\%)\cite{JEM-EUSO}. For measuring EECRs, JEM-EUSO have an advantage over Auger and TA. So determining the EECRs spectrum will depend on the beginning of taking data at JEM-EUSO or the long running time for Auger or TA. This might prove whether there exist SHDM and UHE LFDM in the Universe.
\section{Acknowledgements}
This work was supported by the National Natural Science Foundation
of China (NSFC) under the contract No. 11235006, the Science Fund of
Fujian University of Technology under the contract No. GY-Z14061 and the Natural Science Foundation of
Fujian Province in China under the contract No. 2015J01577.
\par

\newpage

\begin{figure}
 \centering
 \includegraphics[width=0.9\textwidth]{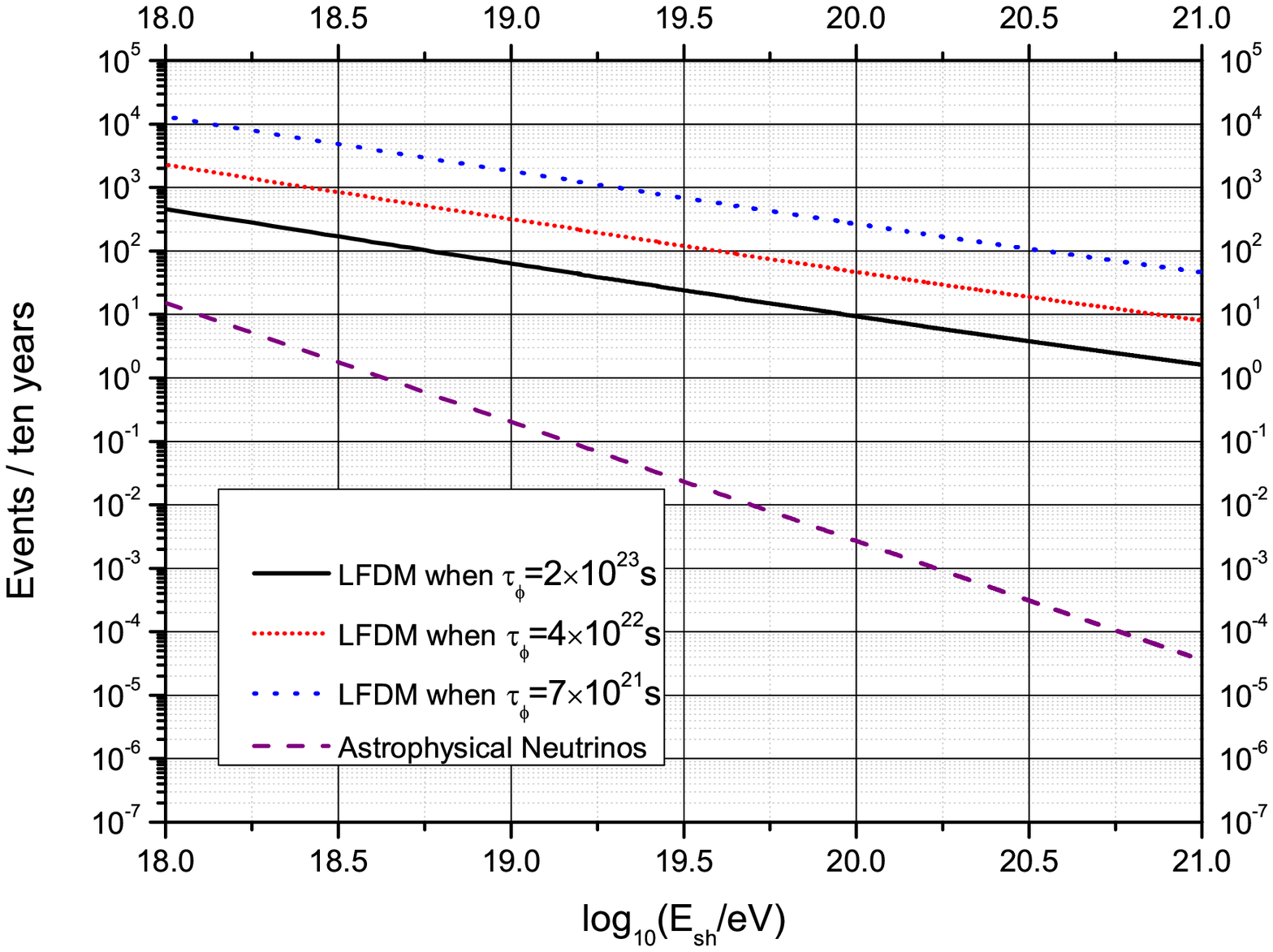}
 \caption{The UHE LFDM and astrophysical neutrino event rates are evaluated at Auger. The red dash dot line is denoting the LFDM event rate when $\tau_{\phi}=1.5\times10^{21}$ s. The blue dot line is denoting the LFDM event rate when $\tau_{\phi}=7\times10^{21}$ s. The black short dot line is denoting the LFDM event rate when $\tau_{\phi}=5\times10^{22}$ s. The black solid line is denoting the LFDM event rate when $\tau_{\phi}=2\times10^{23}$ s. The purple dash line is denoting the astrophysical neutrino event rate.}
 \label{fig:event_rate}
\end{figure}

\begin{figure}
 \centering
 \includegraphics[width=0.9\textwidth]{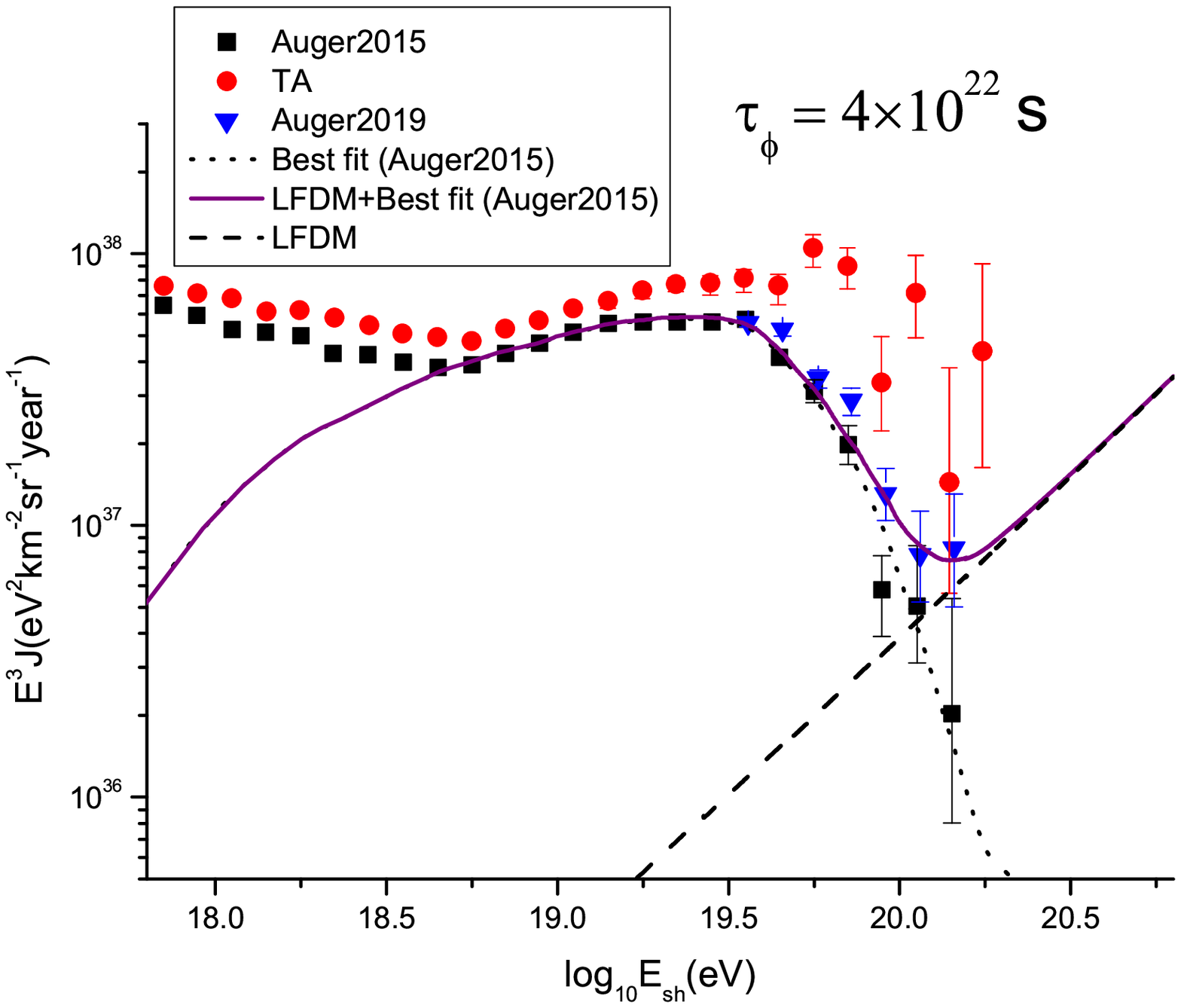}
 \caption{The EECRs spectrum: LFDM spectrum when $\tau_{\phi} = 4\times10^{22}$ s (black dash line), the EECRs data measured by Auger, including the ICRC 2015 Auger data\cite{Auger2015} (full black square) and ICRC 2019 Auger data\cite{Auger2019} (blue triangle), TA (full red circle) and the total (purple solid line) of the LFDM spectrum and ICRC 2015 Auger best-fit spectrum (black dot line)}
 \label{fig:flux_4e22}
\end{figure}

\begin{figure}
 \centering
 \includegraphics[width=0.9\textwidth]{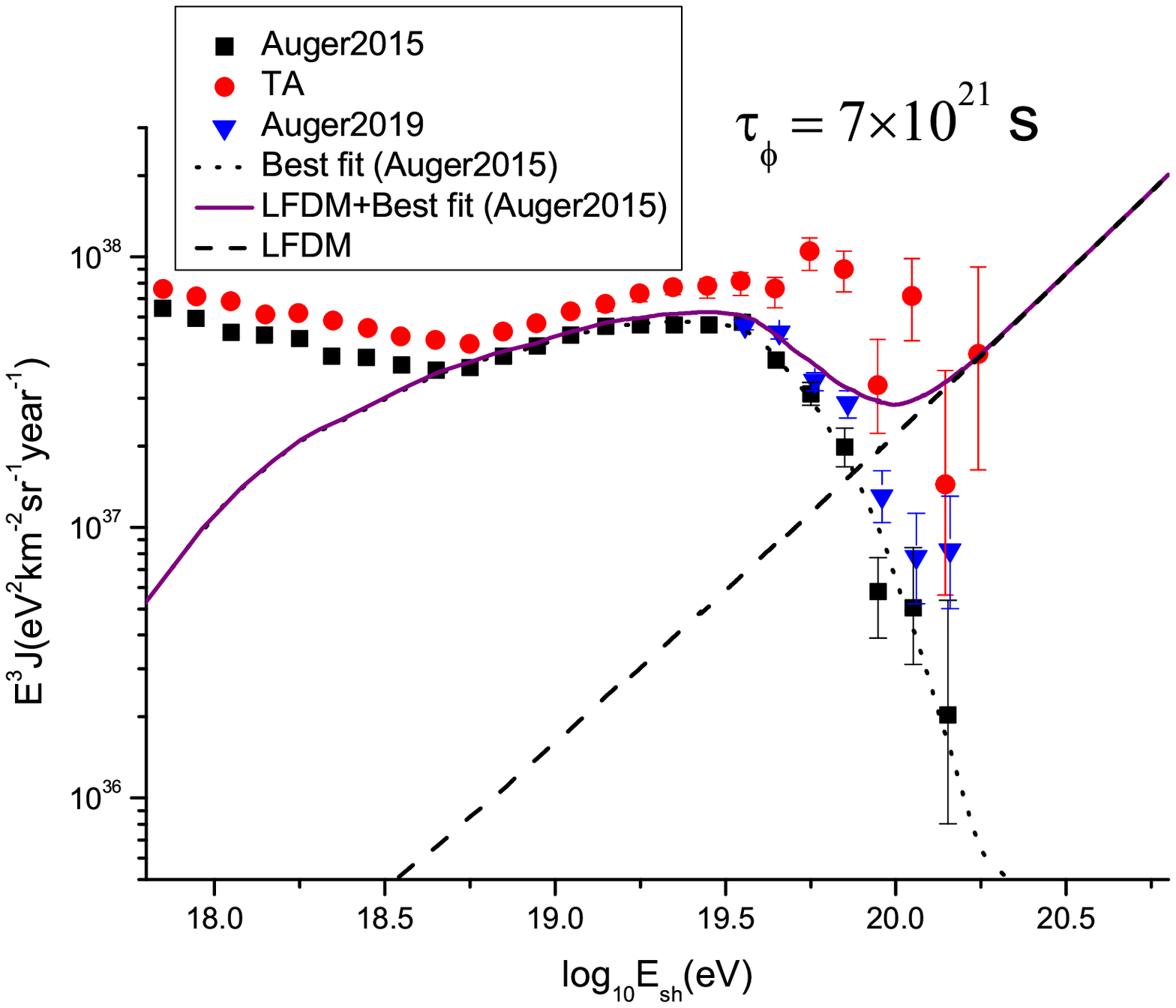}
 \caption{The EECRs spectrum: LFDM spectrum when $\tau_{\phi} = 7\times10^{21}$ s (black dash line), the EECRs data measured by Auger, including the ICRC 2015 Auger data\cite{Auger2015} (full black square) and ICRC 2019 Auger data\cite{Auger2019} (blue triangle), TA (full red circle) and the total (purple solid line) of the LFDM spectrum and ICRC 2015 Auger best-fit spectrum (black dot line)}
 \label{fig:flux_7e21x}
\end{figure}

\end{document}